# Room Temperature Spin to Charge Conversion in Amorphous Topological Insulating Gd-Alloyed $Bi_xSe_{1-x}$/CoFeB Bilayers

Protyush Sahu[1#], Yifei Yang[1#], Yihong Fan[1], Henri Jaffrès[3], Jun-Yang Chen[2], Xavier Devaux[4], Yannick Fagot-Revurat[4], Sylvie Migot[4], Enzo Rongione[3], Tongxin Chen[4], Pambiang Abel Dainone[4], Jean-Marie George[3], Sukhdeep Dhillon[5], Martin Micica[5], Yuan Lu[4]*, and Jian-Ping Wang[1,2⁑]

[1]*School of Physics and Astronomy, University of Minnesota, 116 Church Street SE, Minneapolis, MN 55455, USA*
[2]*Department of Electrical and Computer Engineering, University of Minnesota, 200 Union Street SE, Minneapolis, MN 55455, USA*
[3]*Unité Mixte de Physique, CNRS, Thales, Université Paris-Saclay, 91767, Palaiseau, France*
[4]*Université de Lorraine, CNRS, Institut Jean Lamour, UMR 7198, campus ARTEM, 2 Allée André Guinier, 54011 Nancy, France*
[5]*Laboratoire de Physique de l'Ecole Normale Supérieure, ENS, Université PSL, CNRS, Sorbonne Université, Université Paris Cité, F-75005 Paris, France*

# authors having equal contributions

corresponding authors: *yuan.lu@univ-lorraine.fr; ⁑jpwang@umn.edu

## Abstract

Disordered topological insulator (TI) films have gained intense interest by benefiting from both the TI's exotic transport properties and the advantage of mass production by sputtering. Here, we report on the clear evidence of spin-charge conversion (SCC) in amorphous Gd-alloyed $Bi_xSe_{1-x}$ (BSG)/CoFeB bilayers fabricated by sputtering, which could be related to the amorphous TI surface states. Two methods have been employed to study SCC in BSG($t_{BSG}$=6-16 nm)/CoFeB(5 nm) bilayers with different BSG thicknesses. Firstly, spin pumping is used to generate a spin current in CoFeB and to detect SCC by inverse Edelstein effect (IEE). The maximum SCC efficiency (SCE) is measured as large as 0.035 nm (IEE length $\lambda_{IEE}$) in a 6 nm thick BSG sample, which shows a strong decay when $t_{BSG}$ increases due to the increase of BSG surface roughness. The second method is the THz time-domain spectroscopy, which reveals a small $t_{BSG}$ dependence of SCE, validating the occurrence of a pure interface state related SCC. Furthermore, our angle-resolved photoemission spectroscopy data show dispersive two-dimensional surface states that cross the bulk gap until to the Fermi level, strengthening the possibility of SCC due to the amorphous TI states. Our studies provide a new experimental direction towards the search for topological systems in the amorphous solids.

## Introduction

Bismuth selenide ($Bi_2Se_3$) based topologic insulator (TI) materials have recently received significant attention in the field of condensed matter physics due to their exotic transport properties. Recent interests have garnered from the fact that these materials can exhibit topologically protected surface states[1,2,3,4,5] protected by time reversal symmetry, making them extremely important for various spintronic applications.[6,7,8,9] Many works have demonstrated efficient spin-to-charge interconversion in TI thin films with the aim of magnetization manipulation.[10,11,12,13,14] However, the film growth still remains a major bottleneck since molecular beam epitaxy (MBE) is usually required to grow highly lattice ordered films on the appropriate substrate.[15,16,17] Recently, the presence of strong spin-orbit coupling (SOC) in sputtered polycrystalline $Bi_xSe_{1-x}$ have been demonstrated.[18,19,20,21,22] Quantum transport simulations suggest that the high SOC in sputtered $Bi_xSe_{1-x}$ is due to the quantum confinement effect with an enhanced charge-to-spin conversion efficiency by reduced size and dimensionality,[19] which is different from the typical heavy metal/ferromagnetic bilayer system. These discoveries reveal novel physics in disordered TI materials and shed light on the potential for developing industrial applications. The absence of a well-defined crystal lattice in disordered TI makes them less susceptible to scattering and other perturbations, potentially leading to improved performance in the presence of impurities or imperfections compared to the crystalline TI.

In most of TIs with ordered crystallographic structures, the spin-charge conversion (SCC) is a result of the inverse Edelstein effect (IEE) by the TI surface states.[23,24] For highly disordered TI films, there are two possibilities for the SCC. Firstly, studies have shown that the bulk of TI with low bulk resistivity may play an important role for the SCC by conventional inverse spin Hall effect (ISHE).[25,26] Secondly, theoretical analyses have also shown the possibility of the existence of conducting surface states in amorphous TI films.[27,28,29,30,31,32,33] Spin-orbit torque effects have been observed in different amorphous materials experimentally since 2019.[34,35,36,37] Recently, Corbae *et al.* have evidenced the TI surface states in amorphous $Bi_xSe_{1-x}$ film by angle-resolved photoemission spectroscopy (ARPES).[38] Costa *et al.* also presented an experimental study of the electronic and transport properties of amorphous bismuthene systems,[39] showing that these materials may present topological properties. Moreover, the effects of various dopants in TIs have also been extensively studied. One important advantage of doping is that by embedding atoms with magnetic moment inside the alloys, one can create magnetic topological insulators with time reversal symmetry breaking to observe the quantum anomalous Hall effect (QAHE).[40,41] Incorporating such impurities in TI materials could also introduce highly disordered or even amorphous structures. Therefore, it will be of great interest to explore the role of amorphous TI surface states on the spin-to-charge conversion efficiency for developing scalable spintronic devices.

To study SCC, transport techniques, such as spin-torque ferromagnetic resonance (ST-FMR)[42,43,44,45,46] and spin pumping (SP) combined with the inverse spin-Hall effect,[47,48,49] are often used to measure the conversion efficiency between the charge current and the spin current or spin accumulation in ferromagnetic/non-magnetic (FM/NM) bilayer heterostructures. On the other hand, besides transport techniques, terahertz time-domain spectroscopy (THz-TDS) has also recently emerged as a powerful tool to study the interplay between charge and spin degrees of freedom by all-optical means.[50,51,52,53] The observed terahertz signal originates from the transient charge current converted from a photon-generated spin current into magnetic heterostructures. Compared to ST-FMR and spin pumping, THz-TDS possesses the features of a high time resolution driven by an ultrafast out of equilibrium spin-current that complement the study of SCC via those transport methods.

In this work, we have synthesized Gd-alloyed $Bi_xSe_{1-x}$ (BSG) thin films by magnetron sputtering. Structural and chemical characterizations display a mainly amorphous feature in the disordered BSG film. We have employed both SP-ISHE/IEE transport and THz-TDS optical methods to study the SCC in BSG ($t_{BSG}$=6, 8, 12, 16 nm)/CoFeB(5 nm) bilayer structures. Both methods clearly evidence SCC signatures occurring at BSG/CoFeB interface. Combining the results from ARPES, it is reasonable to attribute the SCC to the amorphous TI surface states at BSG/CoFeB interface. Our study could open a way to discover amorphous TI for the development of scalable topological spintronic device.

## Results and discussions

*Sample preparation*

The multilayer stacks of MgO(2 nm)/Bi(20%)-Se(40%)-Gd(40%)(BSG)/$Co_{20}Fe_{60}B_{20}$(CFB)(5 nm)/MgO(2 nm)/Ta(2 nm) were grown by DC magnetron sputtering (with a base pressure $7\times10^{-8}$ torr) on thermally oxidized silicon substrates. Samples with different thicknesses of BSG ($t_{BSG}$=6, 8, 12, 16 nm) were prepared, which are labeled as BSG6, BSG8, BSG12 and BSG16, respectively. After the growth, the samples were processed by UV lithography for spin pumping measurements. The full film samples were also used for THz-TDS measurements. In addition, we have prepared a bare $SiO_2$//MgO(2 nm)/BSG(8 nm) sample for ARPES characterizations. The readers can find details in the **Methods** section.

*Interfacial structure and chemical properties*

High-resolution transmission electron microscopy (HRTEM) combined with electron energy loss spectroscopy (EELS) and energy-dispersive spectroscopy (EDS) were used for the interfacial structure and chemical characterizations on the BSG8 and BSG16 samples with Ta(5 nm) capping layer. **Fig. 1(a) and (b)** shows the large-scale scanning transmission electron microscopy-high angle annular dark field image (STEM-HAADF) and magnified HRTEM image of the BSG8 sample, respectively. The sample shows relatively flat and sharp interfaces in the large-scale image (**Fig. 1(a)**). In **Fig. 1(b)**, different layers can be distinguished. Since no annealing has been performed on the sample, the CFB layer retains an amorphous feature.[54] The BSG layer also exhibits a highly disordered amorphous character, as evidenced by FFT diffraction pattern (inset of **Fig. 1(b)**) on the blue dashed square zone in **Fig. 1(b)**. The bottom MgO barrier exhibits textured features characterized by small nano-crystallites, while the top MgO layer appears mainly amorphous. On the other hand, the interface chemical distribution has been characterized by EELS. **Fig. 1(c)** displays the different element maps drawn by EELS spectrum images on the region indicated by the red dashed rectangle zone marked in **Fig. 1(a)**. **Fig. 1(d)** displays the element profile and each data point presents an average of element intensity in a zone of 10 nm width and 0.5 nm depth. From the chemical maps and profiles, several pieces of information can be drawn. The distribution of Gd and Se is not homogenous in the BSG layer. Gd has a tendency to accumulate towards the bottom while, on the contrary, Se tends to segregate to the interface with CFB. This could be due to the different enthalpy of formation with bottom MgO layer for Gd and Se elements. Similar variation of the elemental composition is also observed in GdFeCo film grown by co-sputtering.[55] The mean atomic concentration of this area was evaluated as Bi 20%:Se 40%:Gd 40%. The ratio between Co and Fe in the CFB layer was estimated to be 1:4 without considering the B concentration (not acquired).

**Fig. 2(a)** shows the HRTEM image of the sample BSG16. It is found, here, that the interface roughness becomes important. By the guide of black dashed lines shown in **Fig. 2(a)**, it is evident that the increase of interface roughness is mainly attributed to the BSG layer. CFB, MgO and Ta layers well follow the morphology of the BSG layer. Some nanocrystals inside the BSG layer can be evidenced, as marked by the red dashed zones. Energy dispersive spectroscopy (EDS) element mapping images are shown in **Fig. 2(c-e)** with the corresponding STEM-HAADF image in **Fig. 2(b)**. Many zones with brighter contrast can be observed inside the BSG layer in **Fig. 2(b)**, which is due to an inhomogeneous chemical distribution and points out the segregation of some elements. From the chemical maps, the bright contrast zone (marked with red arrow at BSG/CFB interface) is identified to be Bi rich with less Se and Gd, which can also be correlated to the nanocrystals observed in **Fig. 2(a)**. Therefore, during the growth of a thick BSG layer, the enhanced segregation of chemical elements forms the Bi-rich nanocrystals, which results in a rough interface for the BSG16 sample.

*Spin pumping measurements*

We now turn to spin pumping experiments to evaluate the spin-to-charge conversion efficiency in CFB(5 nm)/BSG($t$=6,8,12,16 nm) samples. **Figs. 3(a)** and **(b)** show the schematics of spin pumping measurement and the surface of the device, respectively. A symmetric GSG waveguide is patterned on the top of BSG/CFB structure, which is excited by GHz current to introduce a RF magnetic field in the CFB film. At the condition of ferromagnetic resonance, due to the spin pumping effect, a net spin current ($J_s$) is injected into the BSG layer and afterward converted into a charge current ($J_c$) due to the spin-to-charge conversion process. This gives rise to the generation of a DC voltage (electromotive force (EMF) voltage) across the device. The inset of **Fig. 3(c)** shows a typical spectrum of the EMF voltage acquired at the resonance frequency (9 GHz) for the BSG6 sample. The EMF spectrum is composed of two parts,[47,56] a symmetric Lorentzian shaped signal plus an antisymmetric signal, which can be decomposed into $V(H) = V_{\text{offset}} + V_{\text{sym}} \frac{\Delta H^2}{(H-H_{\text{res}})^2+\Delta H^2} - V_{\text{asym}} \frac{2\Delta H(H-H_{\text{res}})}{(H-H_{\text{res}})^2+\Delta H^2}$. $V_{\text{sym}}$ and $V_{\text{asym}}$ are the magnitude of the symmetric and antisymmetric voltage contributions, respectively. $H_{\text{res}}$ is the resonant field, and $\Delta H$ is the linewidth of the EMF peak. In our case, the EMF signal is characterized by a rather large symmetric contribution which witnesses a clear SCC contribution. In **Fig. 3(c)**, we have shown the variation of $V_{\text{sym}}$ and $V_{\text{asym}}$ as a function of BSG thickness. It is quite clear that the $V_{\text{sym}}$ decreases rapidly with the increase of BSG while $V_{\text{asym}}$ displays a smaller variation for the different thicknesses.

The observed EMF could be attributed to two origins. One is the SP-ISHE in the 3D system where bulk BSG contributes to the spin-to-charge conversion via ISHE. The other one is the SCC in a 2D system at BSG/CFB interface via SP-IEE, as recently evidenced in the case of Rashba interface states[57] and topological insulator surface states[58]. Although the underlying physics of the spin-to-charge conversion is different, SP-IEE has the same key features as SP-ISHE for metallic FM layers.[56] It is reported that ISHE or IEE signals are mostly involved in the symmetric component of EMF signal, while other spin rectification effects may also come into play to mix within the symmetric component of EMF,[49,59,60] such as the interplay between the stray RF currents and the oscillating magnet, related to either the anisotropic magnetoresistance (AMR) or the planar or anomalous Hall effect (PHE or AHE). However, each process is characterized by its own angular dependent signature. Due to the slight asymmetry of our GSG waveguide (see details in **SI Note 2**), the RF excitation field $h_{\mathrm{RF}}$ could have two different contributions: one is in plane along the waveguide and the other one is perpendicular to the layers,[61] although the contribution of the in-plane is expected to be much larger than that of the out-of-plane.[34] As reported in Ref.[59], the contribution of the SCC in the symmetric voltage displays the following angular dependence: $V_{sym,SCC}^{h_ip} = A_{SCC}\frac{\omega_\phi}{\omega(\omega_\theta+\omega_\phi)^2}\cos^3\phi_0$ for the in-plane excitation RF field contribution and $V_{sym,SCC}^{h_oop} = A_{SCC}\frac{\omega_\phi}{\omega(\omega_\theta+\omega_\phi)^2}\cos\phi_0$ for the out-of-plane excitation contribution, where $\omega$, $\omega_\theta$ and $\omega_\phi$ are the angular frequency of the magnetization precession at resonance and its respective components in polar coordinates. In contrast, the AMR and AHE-like signals are characterized by the following angular dependence: $V_{sym,rect}^{h_ip} = A_{rect}\cos\phi_0\frac{\omega_\phi\Delta\rho_{\mathrm{AMR}}\left(\cos 2\ {}_0\mathrm{Im}\left[j_{rf}^{x*}\right]+\sin 2\phi_0\mathrm{Im}\left[j_{rf}^{y*}\right]\right)+\omega\rho_{AHE}\mathrm{Re}[j_{rf}^{x*}]}{\omega(\omega_\theta+\omega_\phi)}$ for the in-plane RF excitation, and $V_{sym,rect}^{h_oop} = A_{rect}\frac{\omega\Delta\rho_{\mathrm{AMR}}\left(\cos 2\phi_0\mathrm{Re}\left[j_{rf}^{x*}\right]+\sin 2\ {}_0\mathrm{Re}\left[j_{rf}^{y*}\right]\right)-\omega_\theta\rho_{AHE}\mathrm{Im}[j_{rf}^{x*}]}{\omega(\omega_\theta+\omega_\phi)}$ for the out-of-plane RF excitation, where $\mathrm{Im}\left[j_{rf}^{y*}\right]$ and $\mathrm{Re}[j_{rf}^{x*}]$ are respectively the imaginary and real parts of the complex conjugate of the RF current passing through the sample along the respective *y* and *x* directions. $\phi_0$ is the angle between the waveguide direction and the in-plane CFB magnetization direction (approximately equal to the in-plane DC field direction when the magnetization is saturated at a sufficiently large magnetic field, see in the inset of **Fig. 3(b)**). Without giving more details, we have shown in **Fig. 4(a)** and **(b)** the results of the angular fit for two samples (BSG8 and BSG16), respectively, considering both in-plane and out-of-plane RF field excitation. It can be clearly seen that both samples have similar angle dependence and that SCC and AMR signals both coexist in the EMF signals. Although the AMR related out-of-plane RF field contribution keeps almost unchanged, the in-plane field contribution (SCC+AMR) is halved in BSG16 sample compared to BSG8 sample. The main conclusion can be drawn: the signal of EMF at $\phi_0 = 0$ is almost attributed to the SCC contribution.

This angular dependent measurement validates that the variation of EMF *vs.* $t_{BSG}$ shown in **Fig. 3(c)** measured at $\phi_0 = 0$ is mostly related to the variation of SCC signals in BSG samples for different thicknesses.

Since the generation of EMF due to ISHE or IEE relies on both the amount of spin current injected and the efficiency of spin-to-charge conversion, it is important to evaluate the spin current density ($J_s$) injected into BSG. We have firstly extracted the corresponding damping parameters ($\alpha$) and saturation magnetization ($M_s$) for the different samples by using a standard fitting procedure from the data acquired at different excitation RF frequencies (see **SI Note 3**). **Fig. 3(d)** displays the variation of $\alpha$ as a function of the BSG thickness. $\alpha$ increases from $5.3\times10^{-3}$ to $1.3\times10^{-2}$ for BSG thicknesses varying from 6 to 16 nm. All these values obtained are in the range of the one obtained from a CoFeB/Pt control sample ($9.6\times10^{-3}$), but are markedly larger than that of a CoFeB/Al reference sample ($1.83\times10^{-3}$) grown on the $SiO_2$ substrate, free of any spin-current dissipation. The latter constitutes an upper bound for the intrinsic damping $\alpha_0$ of a 5 nm thick CoFeB layer. Those measurements reveal a certain spin-current dissipation in the CoFeB layer scaling with the difference of the damping $\Delta\alpha=\alpha-\alpha_0$ and which may be expressed, by part, via the spin-mixing conductance $G^{\uparrow\downarrow} = \frac{4\pi\Delta\alpha M_{eff} t_{CoFeB}}{g_L \mu_B}$ where $g_L$, $\mu_B$, $M_{eff}$, and $t_{CoFeB}$ are, respectively, Landé factor, Bohr magneton, effective magnetization and CoFeB layer thickness. The spin current injected from the CoFeB layer into BSG can be then determined by the following formula[47]:

$$J_S = \frac{G^{\uparrow\downarrow}\gamma^2 {h_{RF}}^2 \hbar \left\{4\pi M_S\gamma + \sqrt{(4\pi M_S\gamma)^2+(2\omega)^2}\right\}}{8\pi\alpha^2 \sqrt{(4\pi M_S\gamma)^2+(2\omega)^2}]} \qquad (2)$$

Here $\gamma$ is the gyromagnetic ratio and $M_S$ is saturation magnetization. We display the variation of $J_s$ in **Fig. 3(e)**, showing a decreasing tendency when BSG thickness increases.

The SCC generated charge current density can be derived from Ohm's law: $J_c^{3D} = \frac{I_c}{A} = \frac{V_{sym}^{SCC}}{Rwt}$ for the 3D system and $J_c^{2D} = \frac{I_c}{w} = \frac{V_{sym}^{SCC}}{Rw}$ for the 2D system, where $R$ is the stripe resistance, $A$ is the section area, $w$ is the stripe width and $t$ is the thickness of the NM layer where the charge current is generated. To understand the current distribution in BSG/CFB interface, we have measured the resistivity of BSG from one specific $SiO_2$//MgO(2 nm)/BSG(16 nm)/MgO(5 nm) sample. It is found that the BSG resistivity (65 mΩ·cm at RT) is much larger than that of CFB (0.16 mΩ·cm at RT). Therefore, the generated charge current is mainly distributed in the CFB layer. Combing with $J_s$ derived from Eq.(2), the spin-to-charge efficiency (SCE=$J_c/J_s$) can be determined for 3D system (also named as spin Hall angle $\theta_{SH}$) or for 2D system (also named as IEE length $\lambda_{IEE}$). More sophisticated calculation of $\theta_{SH}$ should also take into account the spin diffusion length in BSG, which will be discussed below (see **SI Note 4**). As shown in **Fig. 3(f)**, the maximum $\theta_{SH}$ or $\lambda_{IEE}$ can reach 0.02 or 0.035 nm in BSG6 sample, respectively, however they decrease rapidly on increasing the BSG thickness. Although the $\theta_{SH}$ is well lower than the reported value (0.43) for a high ordered $Bi_2Se_3$/CFB system ascribed to the bulk ISHE effect[25] and $\lambda_{IEE}$ is smaller than the value (0.36 nm) reported for an electron beam-evaporated $Bi_2Se_3$/CFB system demonstrating an interface IEE effect,[62] the evident SCE in our case clearly proves the spin-to-charge conversion in BSG/CFB system. In order to further confirm the SCC originating from the BSG layer, we have prepared a reference sample with the stack structure of $SiO_2$//MgO(2 nm)/CFB(5 nm)/MgO(2 nm)/Ta(2 nm). The fitted results of the EMF voltage at 9 GHz show that the symmetric and antisymmetric components are 5.63 and 8.68 μV, respectively. Compared to BSG6 sample, the symmetric component is greatly reduced, thus validating that the SCC mostly occurs due to the BSG layer. In addition, our results can also exclude the possibility of thermal induced anomalous Nernst effect (ANE),[63] which could contribute to the spin pumping signal. This is because the thermal voltage will not show any appreciable thickness dependence unlike we have observed in **Fig. 3(c)**.

*THz-TDS measurements*

Since for both ISHE and IEE, DC-voltage signals from SCC appear in the same direction as $\vec{J_s} \times \hat{\sigma}$ and they are picked up by the same DC-probe configuration placed on the top of BSG/CFB heterostructure, it appears difficult to distinguish them only from spin pumping measurement although the large resistivity of BSG would not favor ISHE mechanism. For that goal, THz-TDS emission spectroscopy in the time domain has recently emerged as a powerful tool to study the interplay between charge and spin degrees of freedom in an all-optical manner. **Fig. 5(a)** shows the schematics of THz-TDS measurements. An ultrafast laser pulse (~100 fs) at 810 nm (200 mW) was launched on the sample surface with a saturating magnetic field applied in the sample plane (between 10 mT and 0.2 T). The generated THz pulses were collected from the front surface (*i.e.*, reflection geometry). In all measurements, we have not observed particularly strong heating mainly due to the fact that the absorption of the laser power is limited to less than 20% for transition metal thin films.[53] Moreover, THz signals showed a strong reproducibility in time on the same series of samples. **Fig. 5(b)** displays the typical emitted electric field as a function of time for the BSG8 sample at RT. In these experiments, the magnetization of CFB has been saturated in two opposite directions giving rise to opposite phase in the THz emission. The THz signal highlights the transient ultrafast charge current from SCC in the BSG/CFB system as expected. It displays a typical lineshape crossing the origin and representative of the derivative of charge current with $E_{THz} \propto \frac{\partial J_c(t)}{\partial t}$. Since the spin current generation in 3$d$ transition metal is reported to be due to the photon-excited hot-electron spin up and spin down populations,[53] the reversal of lineshape of THz signal in **Fig. 5(b)** indicates a sign change of SCC generated charge current related to the magnetization switching in CFB.

**Fig. 5(c)** displays the ensemble of THz-TDS signals from the series of samples with different BSG thicknesses obtained in the same experimental conditions in comparison with a reference sapphire//W(2 nm)/CoFeB(1.8 nm)/Pt(2 nm)/MgO(2 nm). The signals have been temporally shifted for clarity. It reveals that all BSG samples display almost identical lineshapes and amplitudes. The amplitude of BSG samples is about only 5 times smaller than the reference CoFeB/Pt sample, which has been reported to deliver the highest THz emission efficiency and a broadband emission.[64] It has been reported that the self-emission from the ferromagnetic film could exist in the thick Co layer (30 nm) due to the mechanism of photo-induced demagnetization/remagnetization dynamics, while THz emission from SCC plays the main role in the case of thin Co films (3 nm).[65] The clear and relatively large THz signal observed doubly proves the SCC in the BSG/CFB bilayers.[66]

As reported in Ref. [46] in ST-FMR measurement, with an insertion of a Cu layer between the TI and the ferromagnet, the spin accumulation at the surface states can be separately evaluated owing to the suppression of the exchange coupling between the ferromagnet and the surface states of TI possibly detrimental for the latter. Here, we prepared one specific sample (BSG-Cu): $SiO_2$//MgO(2 nm)/BSG(16 nm)/Cu(5 nm)/$Co_{20}Fe_{60}B_{20}$(5 nm)/MgO(2 nm)/Ta(2 nm) characterized by an insertion of a 5 nm thick Cu between BSG and CFB layers. Since the hot electron spin attenuation length in Cu is longer than 33 nm,[67] the spin current injected from CFB layer should be almost conserved at BSG/Cu interface after crossing the 5nm thick Cu layer. We could expect the same amplitude of THz signal (except for the additional THz absorption in Cu) compared to the sample without Cu insertion if the BSG layer truly plays a critical role in the SCC conversion. As shown in **Fig. 5(d)**, the Cu insertion sample shows a reduced signal by a factor about 3 compared to the BSG16 sample which is mainly explained by the additional THz absorption in Cu. After renormalization of THz signal with a factor of 2.66 obtained by considering the resistivity of the thin Cu layer (~20 μΩ·cm[68]) (see details in **SI Note 5**), we indeed obtain almost the same amplitude of signal as the BSG16 sample (**Fig. 5(d)**). This experiment definitively proves that the origin of THz signal is due to the SCC effect by the BSG layer.

*ARPES measurements*

To understand the origin of SCC effect by BSG layer, we have investigated the valence band (VB) of a $SiO_2$//MgO(2 nm)/BSG(8 nm) sample by using ARPES characterization. All spectra were recorded at room temperature. **Fig. 6** summarizes the resulting band structure with the reconstructed data after subtracting the background. **Fig. 6(a)** displays the ARPES intensity map as function of binding energy and wave vector focusing on the region close to the Fermi level. The corresponding X-ray photoelectron spectroscopy (XPS) study of Bi 4f, Gd 3d, Se 3d and O 1s core levels are presented in the **SI Note 6**. The valence band intensity is shown to fall down drastically approaching the topological gap of BSG, *i.e.* for binding energies smaller than -0.5 eV. Moreover, two parallel strongly dispersive states emerge from the VB until to the Fermi level (with black dashed line as a guide to the eye). **Figs. 6(b)** and **(c)** show the momentum dispersion curves (MDCs) taken at and -0.1 eV and -0.3 eV, respectively. The MDCs unambiguously exhibit the existence of two strongly dispersive states in the gap symmetrically centered at $k_{//}= \pm$ 0.28 Å$^{-1}$ ($\theta$=±7°). As reported very recently[38], topologic surface states can exist in amorphous $Bi_2Se_3$ film since a well-defined nearest neighbor distance should correspond to a good reciprocal length scale *i.e.* a good momentum quantum number $k$. The experimental value of $k$ extracted here (less than 0.3 Å$^{-1}$) is very close to the one observed in amorphous $Bi_2Se_3$.[38] Hence, our ARPES measurements clearly evidence the persistence of topological surface states in the band gap of our BSG layers, which could be the origin of SCC effect.

*Discussions*

One important issue is to understand the different qualitative behavior of SCC between spin-pumping and THz-TDS experiments upon BSG thickness. To clarify the different BSG thickness dependence, we have measured the surface roughness of bare BSG layers with different thicknesses by atomic force microscopy (AFM), as plotted in **Fig. 3(d)** compared to the damping parameters (see more details in **SI Note 7**). The RMS roughness of BSG layer is shown to increase with thickness and follows the same tendency as the damping parameter $\alpha$. Such roughness increase can be understood by the formation of Bi crystallites, as clearly evidenced by TEM characterizations in **Fig. 2(a)**. It results in a subsequent and dominant inhomogeneous resonance ascribed to the “orange peel” magnetic coupling[69] more than due to the spin-escape effects. As a consequence, a reduction of spin pumping efficiency as well as the spin mixing conductance is expected with the increase of BSG roughness, as also demonstrated in Pd/Al/Py system[70] for an equivalent roughness. As a result, the spin current density $J_S \propto \frac{G^{\uparrow\downarrow}}{\alpha^2}$ as well as SCE decreases rapidly with increasing BSG thickness as well as $\alpha$. Compared to spin pumping, THz-TDS measurement does not require any resonant condition for SCC, thus it is not sensitive to the roughness and reflects very fairly the true $t_{BSG}$ dependency of SCE.

The small variation measured in THz-TDS suggests that the efficient SCC is already statured in BSG6 sample, which is in good agreement with the spin pumping measurement. For 3D SP-ISHE, $V_{sym}^{SCC}$ *vs.* thickness of NM layer ($d_N$) follows the relationship of $V_{sym}^{SCC} = I_c R = \theta_{SH} R w (\frac{2e}{\hbar}) \lambda \tanh(\frac{d_N}{2\lambda}) j_s^0$, where $\lambda$ is the spin diffusion length in the NM material.[48] The amplitude of $V_{sym}^{SCC}$ increases with the increase of thickness of NM layer, but it saturates at about a thickness of $3\lambda$. In our measurement, since $V_{sym}^{SCC}$ already reaches its maximum at 6 nm, the spin diffusion length in BSG should be smaller than 2 nm. In addition, the resistivity of BSG (65 mΩ·cm) is about 400 times larger than that of CFB (0.16 mΩ·cm), which also indicates that the SCC generated charge current can only exist at the interface of BSG/CFB but not inside BSG owing to the impedance mismatch and the spin-back flow issues[71] when the resistivity of the NM becomes very large. Therefore, it is reasonable to consider that the observed SCC is due to the 2D IEE at BSG/CFB interface involving a Rashba or a TI state.

From our ARPES measurement, the possibility of a topologically protected surface state in our amorphous BSG films is privileged. These results are in good agreement with that measured by Corbae *et al.* on amorphous $Bi_2Se_3$ thin films.[38] They demonstrated a dispersive two-dimensional surface state that crosses the bulk gap. Spin resolved photoemission spectroscopy shows this state has an anti-symmetric spin texture resembling that of the surface state of crystalline $Bi_2Se_3$. As recently predicted by several theoretical work,[27,28,29,30,31,32,33] the topological spin-momentum locking surface states can exist in an amorphous TI surface. The reason is that in random lattices fermions hop between sites within a finite range. By tuning parameters such as the density of states, the system undergoes a quantum phase transition from a trivial to a topological phase that manifests as quantized conductance in the system boundary.[28] If these surface states do exist on the surface of BSG, our sample with Cu insertion indicates that the deposition of Cu has a minor role in varying the surface states of BSG for the SCC, which is consistent with the ST-FMR measurements on $(Bi_{1-x}Sb_x)_2Te_3$/Cu/NiFe trilayers.[46]

## Conclusions

In summary, we have fabricated amorphous Gd-alloyed $Bi_xSe_{1-x}$ thin films by sputtering method and studied the spin-to-charge conversion in BSG/CFB bilayers. Structural characterizations show a mainly amorphous feature for the 8 nm thick BSG film, while Bi rich crystallites appear inside the 16 nm thick BSG film. Spin pumping and THz-TDS measurements have been employed to study SCC in BSG($t_{BSG}$)/CoFeB(5 nm) bilayers with different BSG thicknesses ($t_{BSG}$=6-16 nm). Both methods have proved the efficient SCC at BSG/CFB interface. The maximum SCE is measured as large as 0.035 nm for $\lambda_{IEE}$ in BSG6 sample. The spin pumping measurements show a fast decay of SCE when $t_{BSG}$ increases due to the increase of BSG surface roughness. However, TDS measurements reveal a small $t_{BSG}$ dependence of SCE, validating a pure interface state related SCC. Furthermore, our APRES data is consistent with a dispersive two-dimensional surface state that crosses the bulk gap to Fermi level, privileging the origin of SCC due to the amorphous TI state. Our studies give a deep insight into the SCC in spin pumping and TDS measurements and provide a new experimental direction to the search for topological systems in amorphous solids to develop scalable topological devices.

## Methods

### *Sample preparation:*

The films were grown by magnetron sputtering on a thermally oxidized silicon substrate. The stacks are: MgO(2 nm)/BSG($t$ nm)/CFB(5 nm)/MgO(2 nm)/Ta(2 nm). BSG was grown by co-sputtering of $Bi_2Se_3$ and Gd targets. The Ar flow during sputtering was 40 sccm and the anode bias was 60 V. The cathode power for BiSe and Gd was 30 W and 40 W, respectively. The BSG deposition rate was 0.7 Å/s.

### *TEM characterization:*

HRTEM and STEM were performed to characterize the interfacial structure using a probe corrected JEOL ARM 200 CF operated at 200 kV. Thin lamella was extracted by focused ion beam (FIB) milling using an FEI Helios Nanolab dual beam. EELS spectrum images were recorded in STEM mode with a Gatan Quantum Imaging filter. In order to correct energy drift and estimate the local thickness, the zero-loss and the core-loss spectra were simultaneously recorded for a dispersion of 1 eV (Dual EELS method). The core loss spectra were registered in the range 670-2700 eV in order to record $Fe_L$, $Co_L$, $Gd_M$, $Mg_K$, $Se_L$, $Ta_M$ and $Bi_M$ edges. $B_K$

and $O_K$ signals with edges respectively near 188 eV and 532 eV were not recorded. The pixel size of the spectrum image was fixed at 0.4 nm for a dwell time of 0.5 s/pixel for the core loss spectra. After energy drift correction, the spectrum images were denoised using a principal component analysis method [72] before quantitative analysis. EDS spectrum images were recorded in STEM mode with a JEOL JED2300T silicon-drift detector. Spectrum images were obtained by integrating 170 frames of 256×256 pixels. The pixel size was fixed at 0.23 nm for a dwell time of 0.2 msec.

*Spin pumping measurement:*

We performed spin pumping experiment to measure the spin-to-charge conversion. The samples were patterned into stripes with a width and length of 620 μm and 1500 μm, respectively, using UV photolithography and ion milling. A 55 nm thick silicon dioxide layer was deposited to insulate the wave guide from the film. The wave guides and contact pads were patterned by UV photolithography, and a Ti(10 nm)/Au(150 nm) electrode was deposited. The waveguide of the spin pumping devices has the signal linewidth of 75 μm, ground linewidth of 225 μm, and separation between the lines of 37.5 μm. The RF current generates an excitation magnetic field, which causes the precession of the magnetization of the CoFeB layer at a GHz frequency. When the frequency of magnetic field matches with the oscillation frequency of the FM layer under a certain resonance field, the spin current generated from the CoFeB layer injects into the BSG layer due to the spin pumping effect. The injected spin current is then converted to a DC charge current, which is probed by measuring the open circuit voltage on the two terminates of the stripe using a Keithley 2182A nanovoltmeter.

*THz spectroscopy measurement:*

The film samples were placed on a mount with a small magnetic field (between 10 mT and 0.2 T) parallel to the spin interface. An ultrafast (~100 fs pulses) Ti:Sapphire oscillator centered at 810 nm was used to photo-excite the spin carriers directly from the front surface. Average power of up to ~200 mW was used with a repetition rate of 80 MHz. The typical laser spot size on the sample was about 200×200 $\mu m^2$. The generated THz pulses were collected from the front surface of the spin emitter (*i.e.*, reflection geometry) using a set of parabolic mirrors of 150 and 75 mm focal length to focus on samples. Standard electrooptic sampling was used to detect the electric field of the THz pulses, using a 500-μm-thick <110> ZnTe crystal. A chopper was placed at the focal point between the second and third parabolic mirror to modulate the THz beam at 6 kHz for lock-in detection. The TDS setup was placed in a dry-air purged chamber (typically 3-5% humidity) to reduce the influence of water absorption on the THz radiation.

*ARPES and XPS measurement:*

The sample has been cleaned by two cycles of $Ar^+$ etching (5 mins at 0.75 keV) by monitoring at the same time the O 1s, Bi 4f, Ga 3d and Se 3d core levels as presented in **SI Note 6**. A high flux MBS UV He lamp has been used to produce photons with incident energy of 21.22 eV. A high resolution Scienta DA30L analyzer operating in angular mode with a Pass Energy of 10 eV has been used to record valence band spectra. The ARPES intensity map $I(E_B, \theta)$ has been transformed in $I(E_B, k_{//})$ using the standard conservation law of $k_{//}$. The energy and angular resolution are respectively better than 10 meV and 0.05 $Å^{-1}$. The Fermi level has been measured on Ag paste used to connect electrically the BSG samples to avoid charging effect. The XPS spectra were recorded with the same analyzer with a Pass Energy of 200 eV using a monochromatized Al-Kα X-ray source with an energy of 1486.6 eV. The overall energy resolution in XPS was better than 300 meV.

**Supporting Information.** Magnetic characterization of BSG film; Origin of out-of-plane excitation RF magnetic field; Extraction of damping parameters; Calculation of spin Hall angle in 3D system; XPS of BSG sample for ARPES measurement; AFM measurements of BSG layer with different thicknesses; Comparison of sputtering grown topological insulator materials.

**Acknowledgements**

We acknowledge the discussion with Prof. Albert Fert. This work was supported in part by C-SPIN, one of six centers of STARnet, and is partly supported by ASCENT, one of six centers of JUMP, a Semiconductor Research Corporation program that is sponsored by MARCO and DARPA. Portions of this work were conducted in the Minnesota Nano Center, which is supported by the National Science Foundation through the National Nano Coordinated Infrastructure Network (NNCI) under Award Number ECCS-1542202. Y.Lu acknowledges the support from the French National Research Agency (ANR) FEOrgSpin (No. ANR-18-CE24-0017-01) and SOTspinLED (No. ANR-22-CE24-0006-01) projects.

## Figures

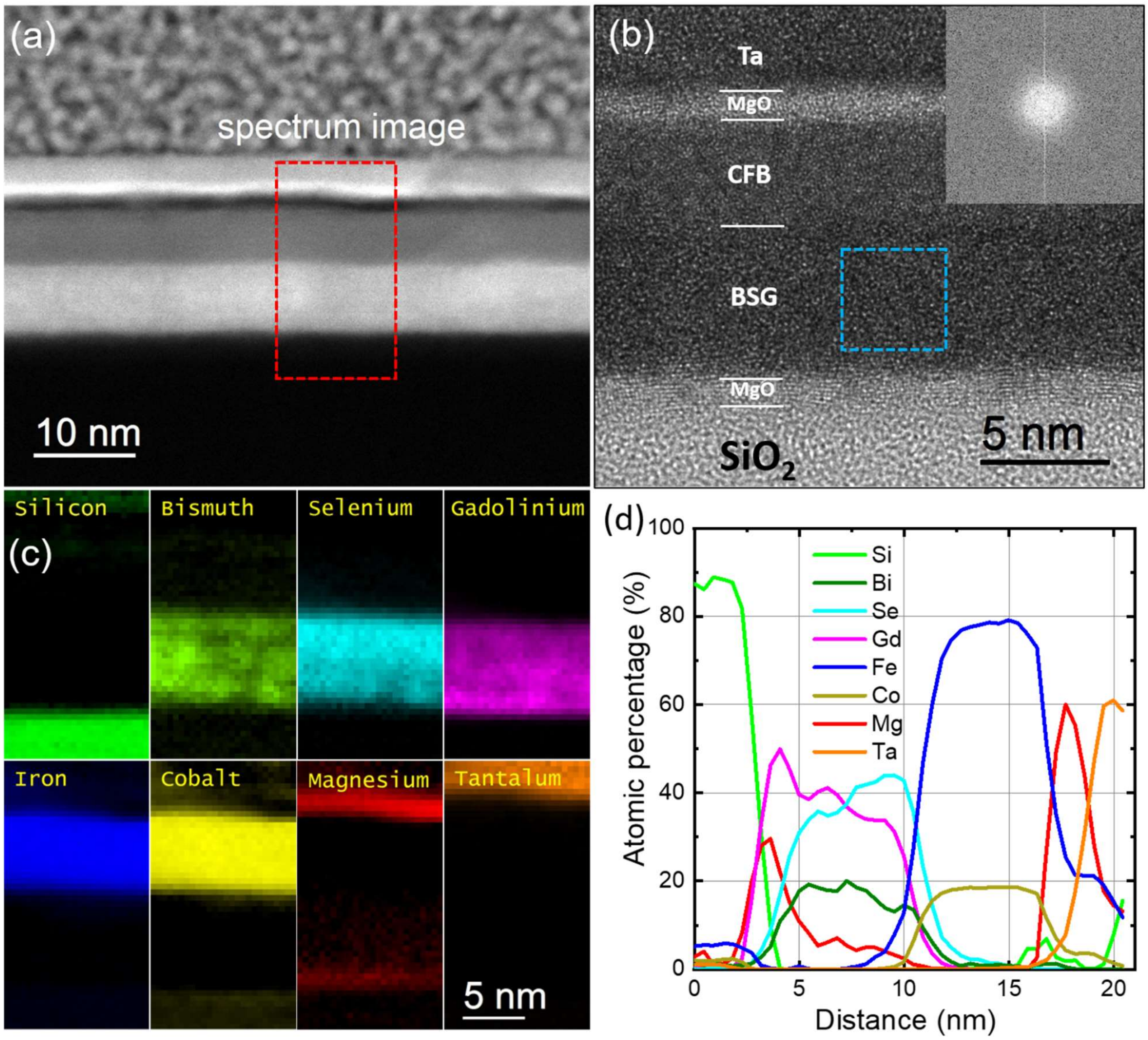

**Figure 1:** Structural and chemical characterization of BSG8 sample by TEM. (a) STEM HAADF image on the BSG8 sample at middle magnification. The red rectangle shows the region where the EELS spectrum images were recorded. (b) HRTEM image in a magnified scale on the BSG8 sample. Inset: FFT pattern on the blue dashed square zone. (c) Elemental maps of the stacks drawn from EELS spectrum images after quantitative processing with all the essential elements from the substrate to the capping layer. The tiny silicon signal visible in the Ta capping layer and the Mg signal inside BSG layer are due to an artefact of processing because of the overlapping of different element edge peaks. (d) Elemental profiles across the multilayer structure drawn from EELS maps.

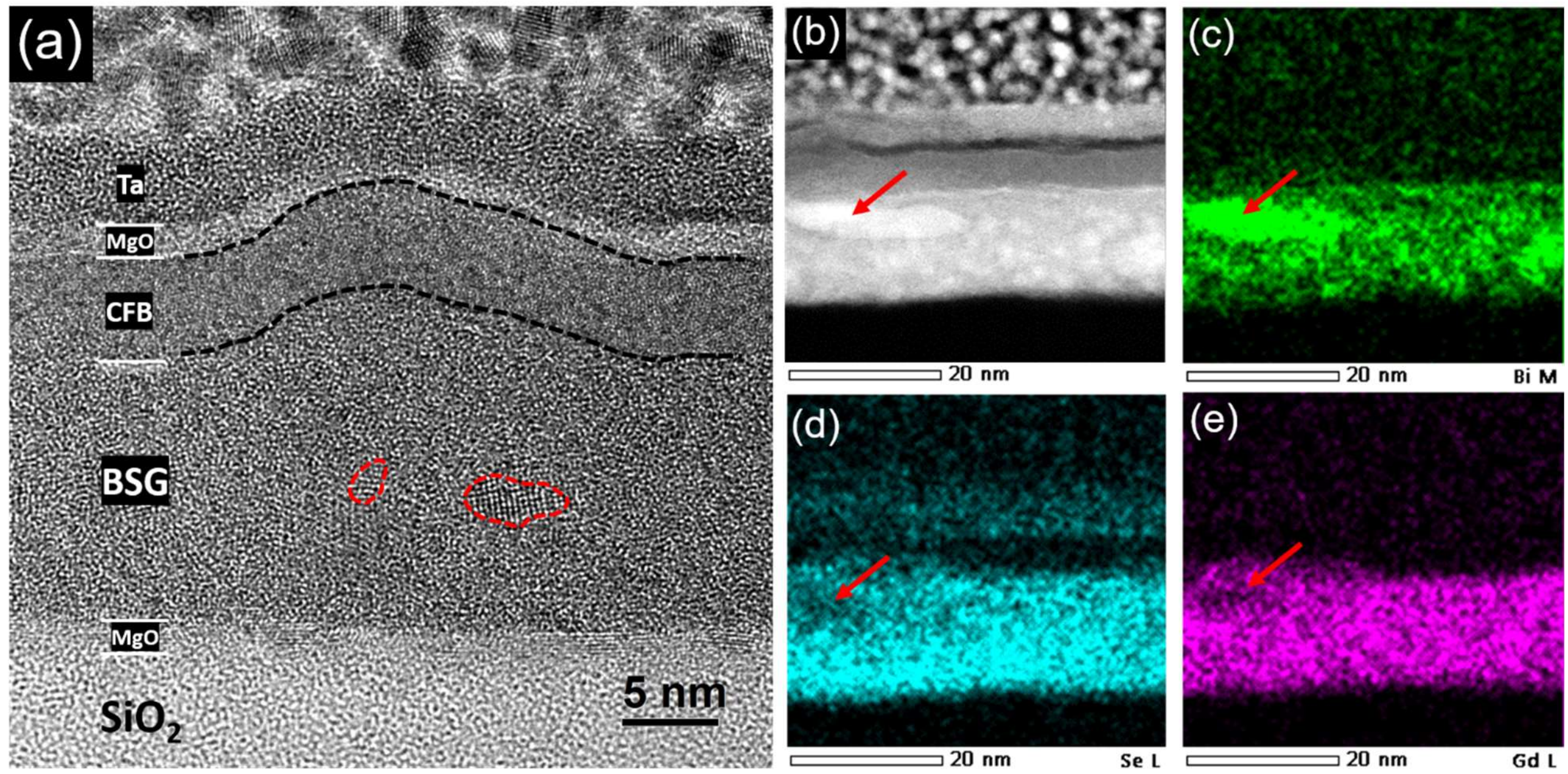

**Figure 2:** Structural and chemical characterization of BSG16 sample by TEM. (a) HRTEM image in magnified scale on the BSG16 sample. The black dashed lines guide the eyes to show the interface roughness due to the BSG layer. The red cycles show the zone where nanocrystals can be found. (b) STEM HAADF image and corresponding EDS elemental maps for (c) Bi, (d) Se and (e) Gd. The red arrows indicate that the white contrast zone in (b) is due to the inhomogeneous element distribution with a Bi-rich character.

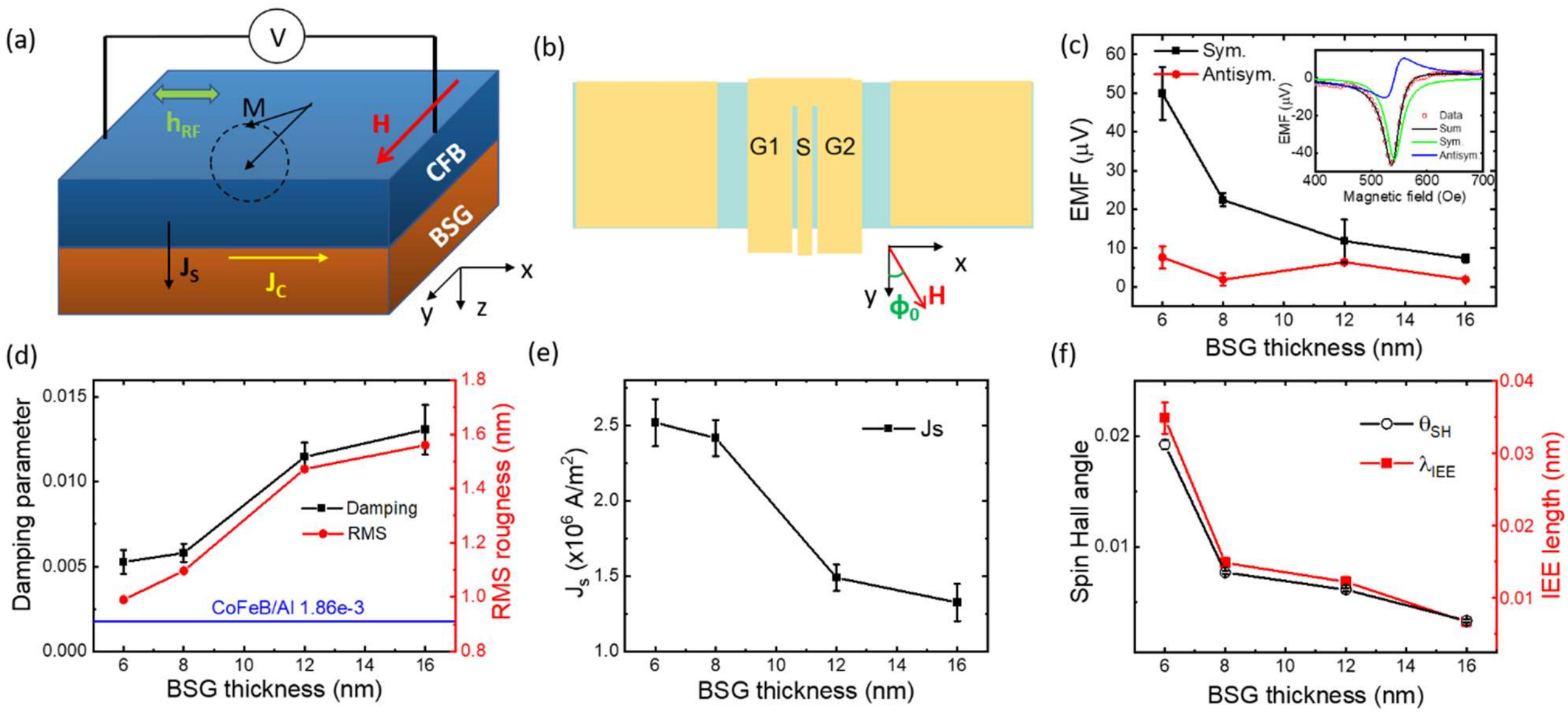

**Figure 3:** Spin-to-charge conversion studied by spin pumping measurements. (a) Schematic of spin pumping measurement. (b) Schematic of the spin pumping device. Inset: schematic configuration of angle dependence measurement. (c) Decomposed symmetric and antisymmetric components of EMF voltage as a function of BSG thickness. Inset: Typical spectra of the EMF voltage acquired at the resonance frequency (9 GHz) for the BSG6 sample. The peak can be fitted by Lorentzian function to decompose the symmetric and antisymmetric components. (d) Damping parameter and BSG root mean square (RMS) roughness as a function of BSG thickness. (e) Spin current density ($J_s$) generated by spin pumping as a function of BSG thickness. (f) Spin-to-charge conversion efficiency (SCE) as a function of BSG thickness presented either as spin Hall angle ($\theta_{SH}$) for 3D system or as IEE length ($\lambda_{IEE}$) for 2D system.

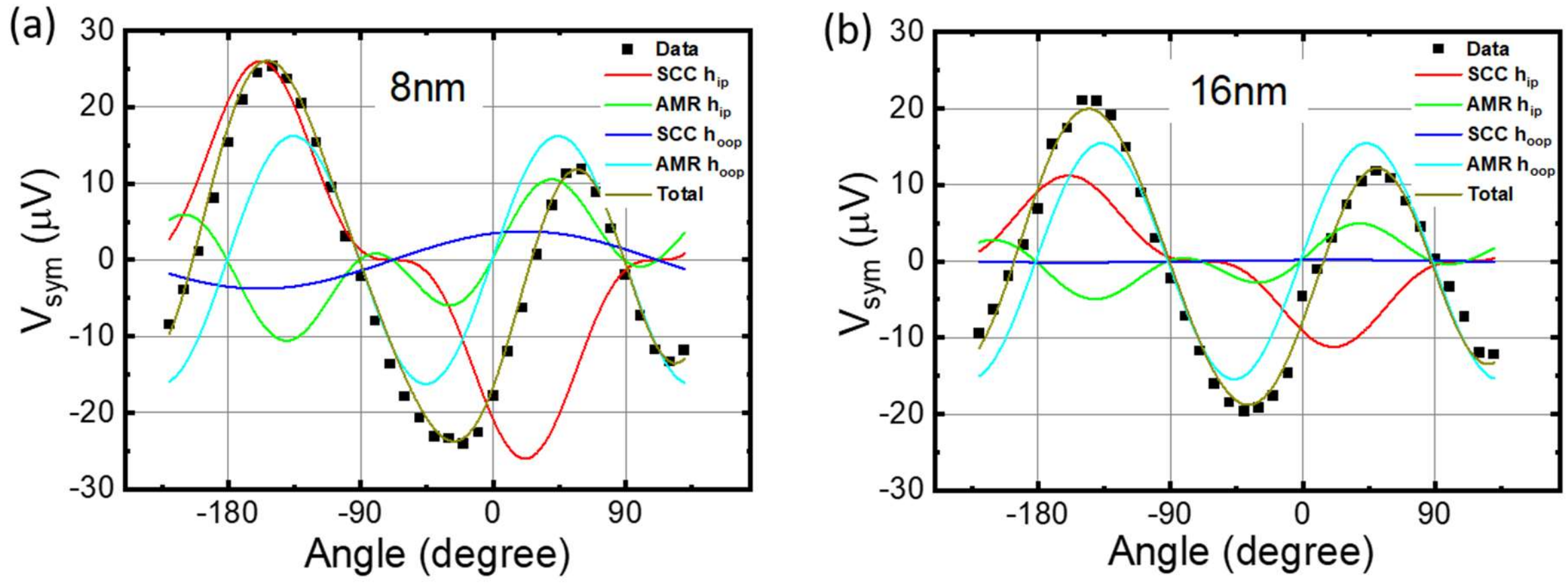

**Figure 4:** Extracted $V_{\text{sym}}$ as a function of the angle $\phi_0$ between the waveguide and the DC magnetic field for BSG/CFB bilayers with (a) 8 nm thick BSG and (b) 16 nm thick BSG. The data was fitted by the model considering SCC and AMR contributions with in-plane and out-of-plane excitation RF magnetic field.

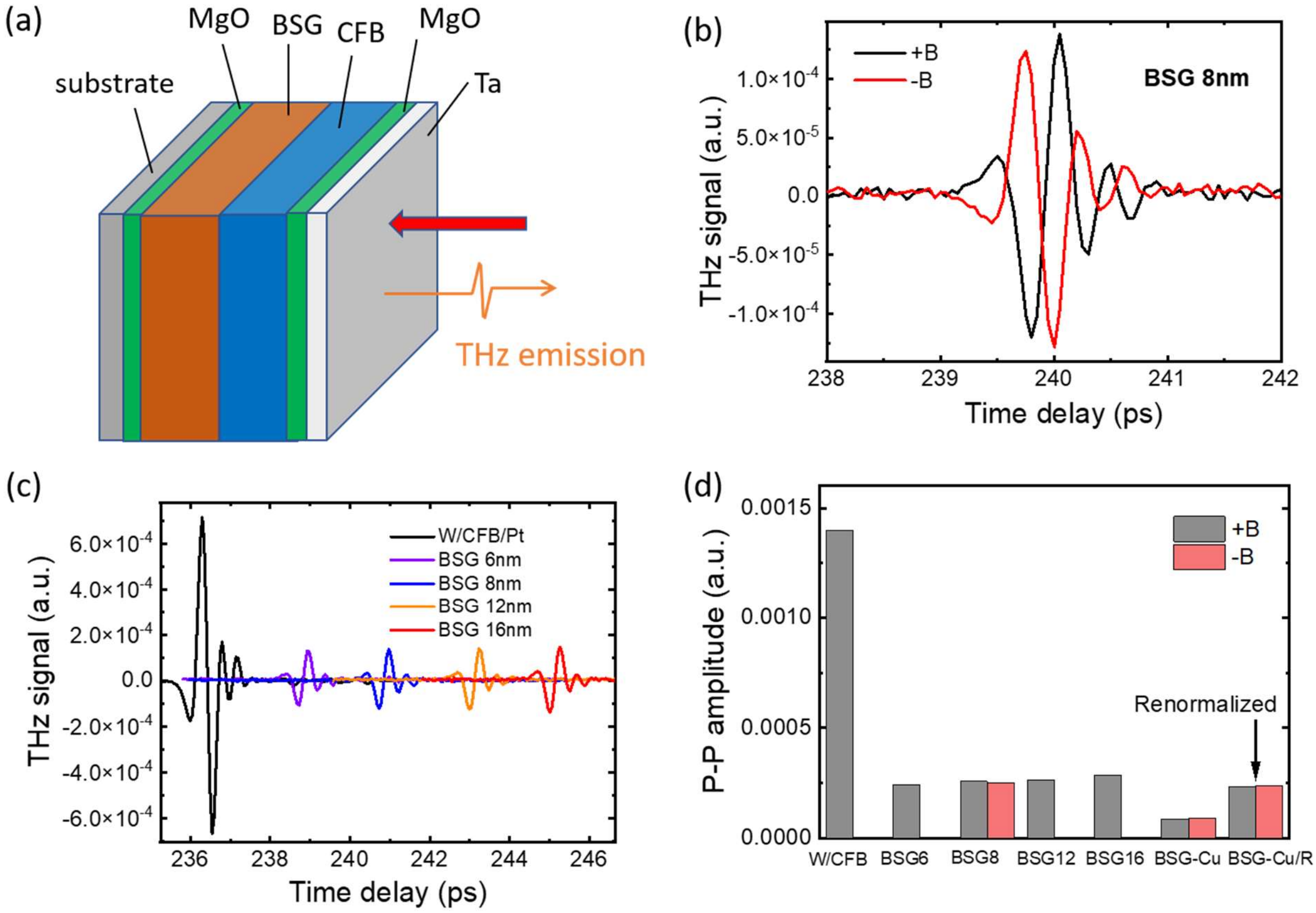

**Figure 5:** Spin-to-charge conversion studied by THz-TDS measurements. (a) Schematic of TDS experimental setup. (b) Typical emitted electric field as a function of time for the BSG8 sample at RT with two opposite applied magnetic fields (±1k Oe). (c) THz-TDS signals from the series of samples with different BSG thicknesses in comparison with a reference sample of sapphire//W(2 nm)/CoFeB(1.8 nm)/Pt(2 nm)/MgO(2 nm). The signals have been temporally shifted for clarity. (d) Peak-to-peak amplitude of THz signal for samples with different BSG thicknesses in comparison with the control sample with an insertion of 5 nm thick Cu between BSG and CFB. The signal of Cu insertion sample has also been renormalized by taking account of THz absorption by the Cu layer.

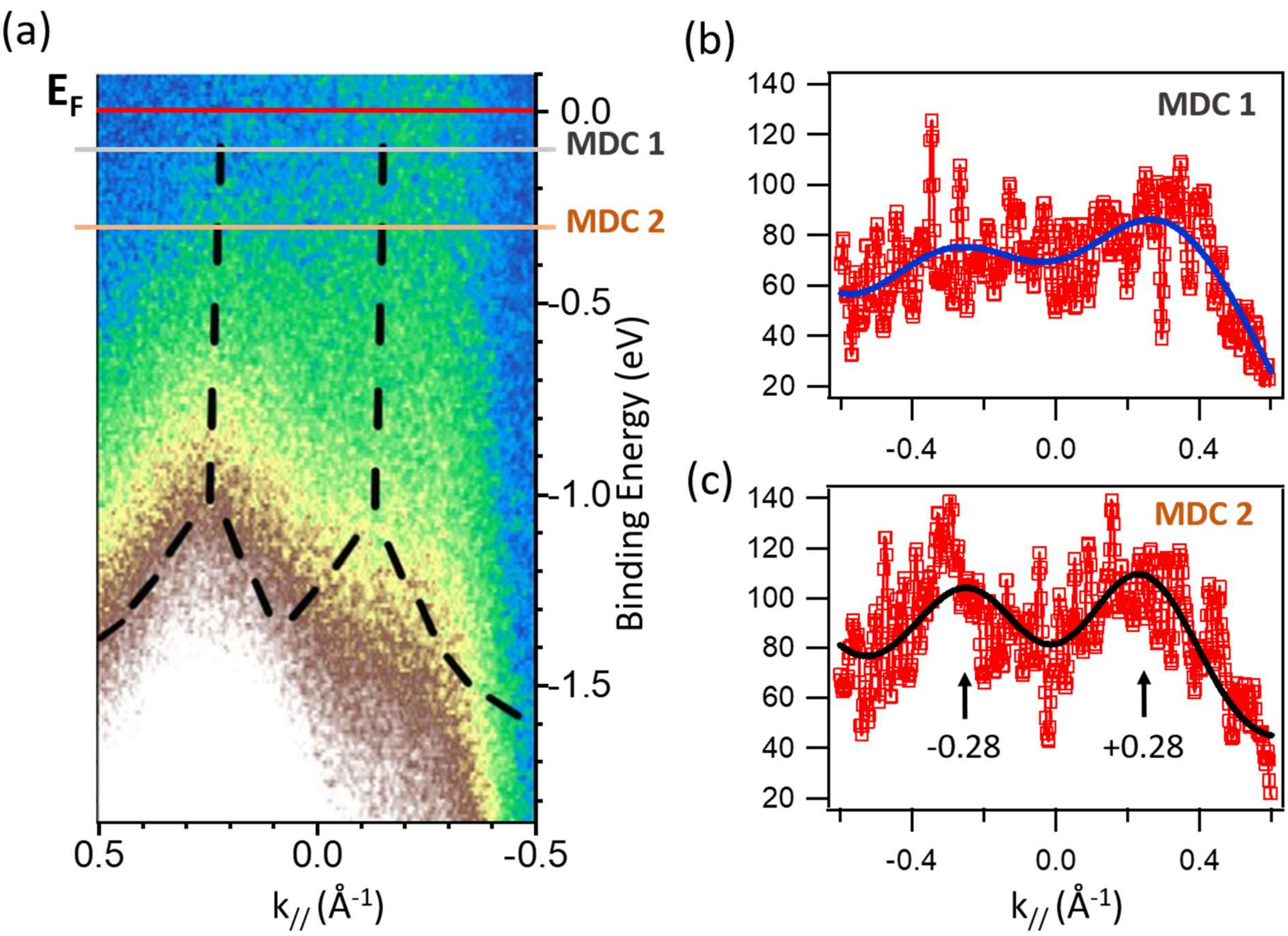

**Figure 6:** Angle-resolved photoemission spectroscopy measurement on $SiO_2$//MgO(2 nm)/BSG(8 nm) sample. (a) ARPES intensity map $I(E_B, k_{//})$. Corresponding Momentum Dispersion Curve (constant energy cut) located in the topological gap of amorphous BSG at (b) -0.1 eV and (c) -0.3 eV.

## TOC figure

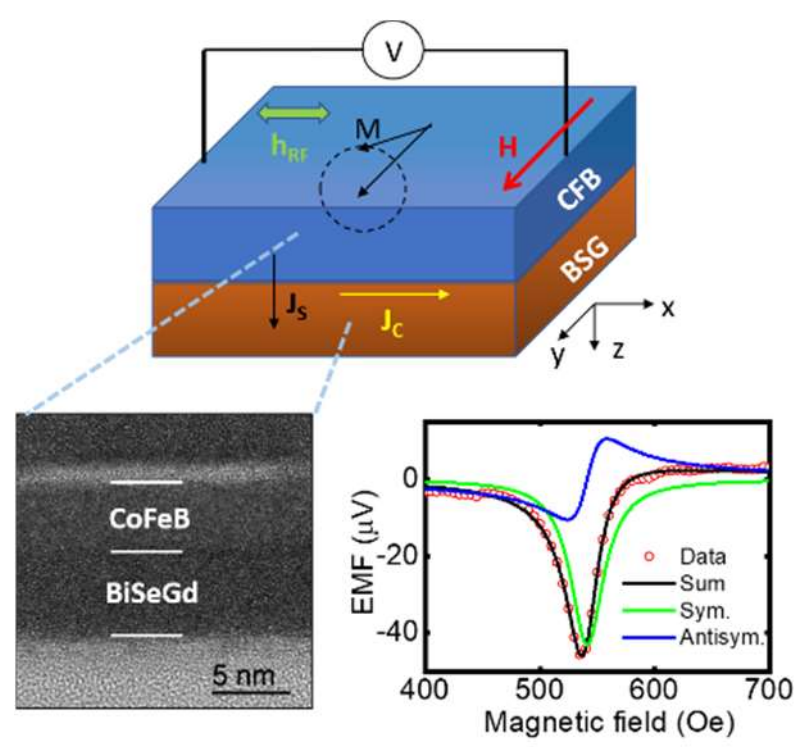

# Supporting Information

## Room Temperature Spin to Charge Conversion in Amorphous Topological Insulating Gd-Alloyed $Bi_xSe_{1-x}$/CoFeB Bilayers

Protyush Sahu[1#], Yifei Yang[1#], Yihong Fan[1], Henri Jaffrès[4], Jun-Yang Chen[2], Xavier Devaux[4], Yannick Fagot-Revurat[4], Sylvie Migot[4], Enzo Rongione[3], Tongxin Chen[4], Pambiang Abel Dainone[4], Jean-Marie George[3], Sukhdeep Dhillon[5], Martin Micica[5], Yuan Lu[4*], and Jian-Ping Wang[1,2⁑]

[1]*School of Physics and Astronomy, University of Minnesota, 116 Church Street SE, Minneapolis, MN 55455, USA*
[2]*Department of Electrical and Computer Engineering, University of Minnesota, 200 Union Street SE, Minneapolis, MN 55455, USA*
[3]*Unité Mixte de Physique, CNRS, Thales, Université Paris-Saclay, 91767, Palaiseau, France*
[4]*Université de Lorraine, CNRS, Institut Jean Lamour, UMR 7198, campus ARTEM, 2 Allée André Guinier, 54011 Nancy, France*
[5]*Laboratoire de Physique de l'Ecole Normale Supérieure, ENS, Université PSL, CNRS, Sorbonne Université, Université Paris Cité, F-75005 Paris, France*

# authors having equal contributions
corresponding authors: *yuan.lu@univ-lorraine.fr; ⁑jpwang@umn.edu

### Note 1. Magnetic characterization of BSG film

We have performed magnetic characterizations by superconducting quantum interference device (SQUID) for the bare 16 nm BSG sample at different temperature, as shown in **Fig. S1**. The curves show only linear variation behavior due to the paramagnetic character of $Si/SiO_2$ substrate. The signal from BSG layer can be extracted after subtracting the linear background. However, we found that the signal from BSG is very small, which indicates that the BSG layer is almost non-magnetic although the Gd atom is magnetic.

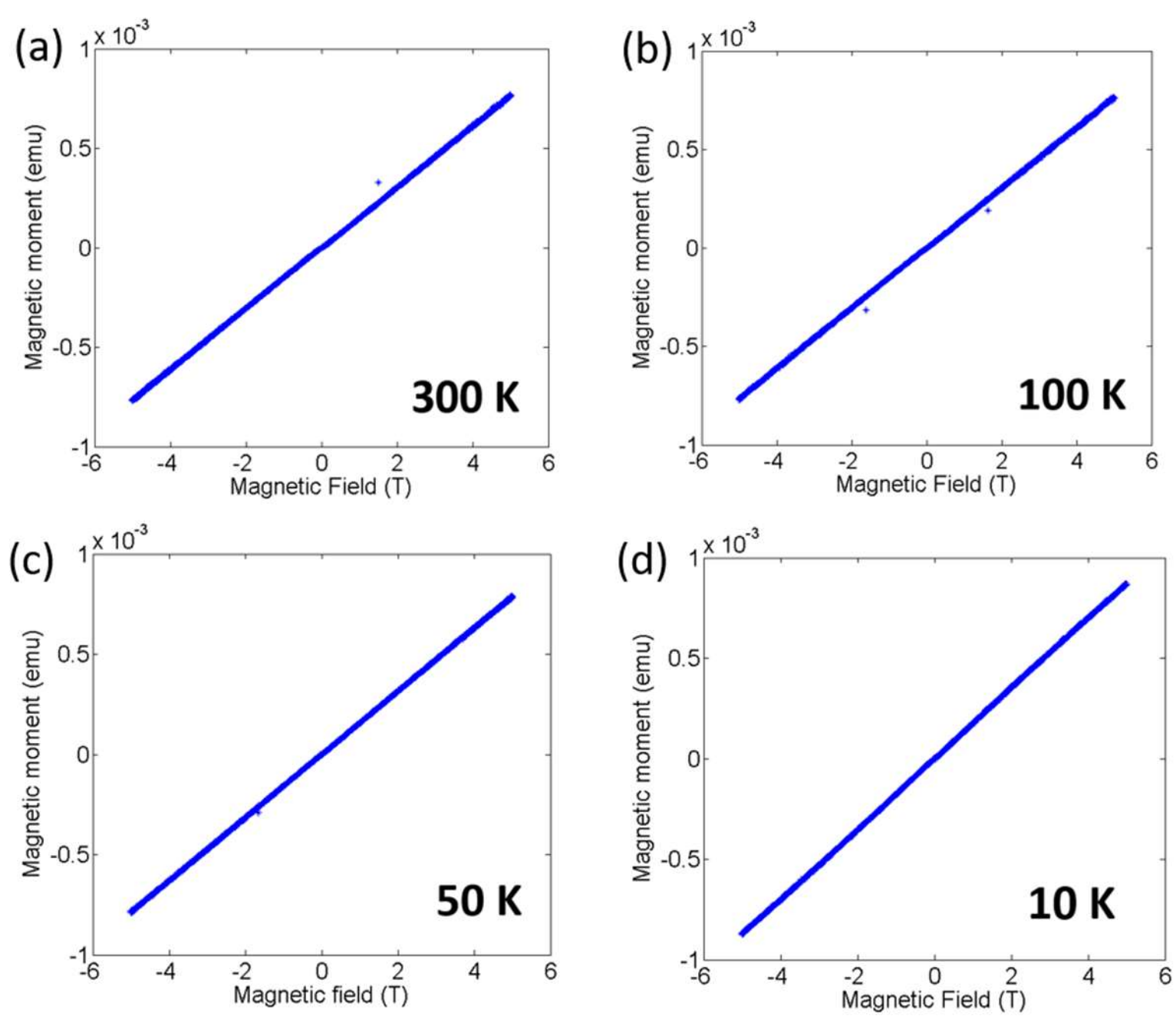

**Fig. S1:** Measurement of *M-H* curves on the bare 16 nm BSG sample at different temperatures: (a) 300 K, (b) 100 K, (c) 50 K and (d) 10 K.

## Note 2. Origin of out-of-plane excitation RF magnetic field

For the spin pumping measurement, we use a GSG waveguide to generate the RF magnetic field, as shown in **Fig. S2(a)**. The waveguide contains one source line (S) and one ground line with two terminals (G1 and G2). Using Ampere's law, the in-plane RF magnetic field generated by the waveguide is estimated to be 0.43 Oe, when an excitation frequency of 9 GHz and an amplitude of 2 V are used. As discussed in the main text, the output voltage of spin pumping has a component associated with the perpendicular RF magnetic field. Here we found that this field can possibly originate from the asymmetric geometry of the spin pumping device.

If the device is perfectly symmetric, as shown in **Fig. S2(b)**, the perpendicular fields induced by source and ground lines will cancel each other out. However, if the device has some asymmetry, which could result from the patterning process, a perpendicular field will arise. **Fig. S2(c)** shows an asymmetric source line, with a width of *d* missing at one side. Since there is more sample area on the left side of the source line, the perpendicular field will be larger than that on the right side. Assuming *d* is 10% of the total width of the source line, the estimated perpendicular magnetic field on the sample is 0.037 Oe, which is about one order of magnitude smaller than the in-plane field.

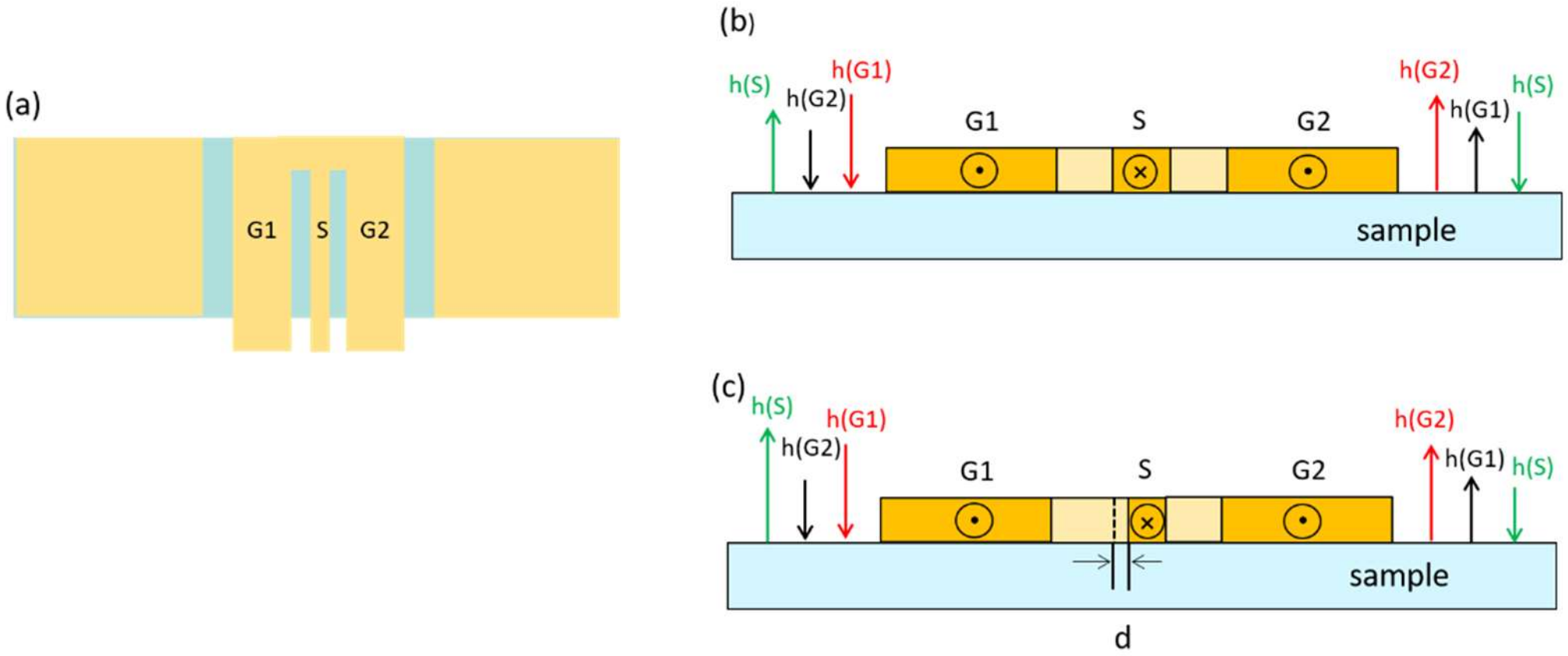

**Fig. S2**: (a) Top view of the spin pumping device. The source line (S) and two ground line terminals (G1 and G2) are shown. (b) Illustration of perpendicular fields generated by a symmetric waveguide. (c) Illustration of perpendicular fields generated by an asymmetric waveguide.

## Note 3. Extraction of damping parameters

To extract the parameters of damping and effective magnetization, we have measured FMR resonant spectra at different resonant frequency from 9 GHz to 17 GHz. As shown in **Fig. S3(a)** for BSG6 sample, with the increase of frequency, the DC magnetic field to obtain FMR also increases. At the meantime, the amplitude of peak decreases and the width of peak increases with increasing the frequency. As shown in **Fig. S3(b)** for 9 GHz excitation, the resonant spectrum can be fitted by a combination of symmetric and anti-symmetric Lorentzian fit following:[1]

$$V = \frac{V_S \Delta H^2}{\Delta H^2 + (H_{ext} - H_{res})^2} + \frac{V_A (H_{ext} - H_{res})}{\Delta H(\Delta H^2 + (H_{ext} - H_{res})^2)} \quad \text{(S1)}$$

$V_S$ is symmetric voltage, $V_A$ is anti-symmetric voltage, $\Delta H$ is linewidth, $H_{res}$ is resonance field, $H_{ext}$ is external DC field. The symmetric voltage is maximized at resonance whereas the anti-symmetric voltage reaches zero at resonance. The effective magnetization ($M_{eff}$) can be fitted by the relationship between the resonant frequency ($f_{res}$) and resonant magnetic field ($H_{res}$) as following:[1]

$$f(H_{res}) = \frac{g_L \mu_B}{h} \sqrt{H_{res}(H_{res} + 4\pi M_{eff})} \quad \text{(S2)}$$

As shown in **Fig.S3(c)**, we get a $M_{eff}$ of 1377±16 kA/m for a the BSG6 sample.

The damping parameter $\alpha$ can be fitted by the relationship between the resonant frequency ($f_{res}$) and peak-peak width ($\Delta H$):

$$\Delta H(f_{res}) = \Delta H_0 + \frac{2}{\sqrt{3}} \alpha \frac{2\pi f_{res}}{g_L \mu_B} \quad \text{(S3)}$$

As shown in **Fig. S3(d)**, the damping extracted from the linear fitting give a value of 0.0053 ±0.0007.

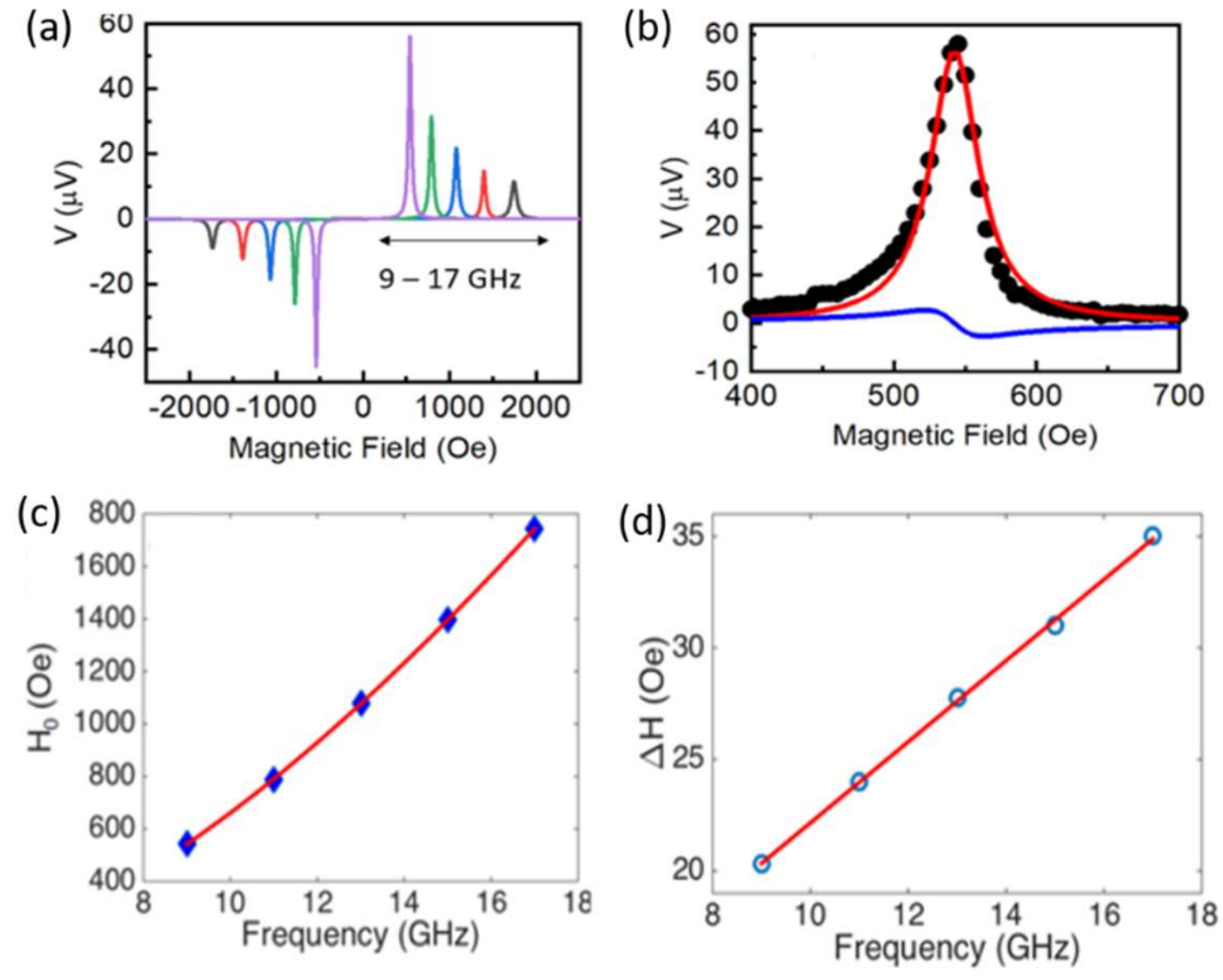

**Fig. S3:** (a) EMF spectra with different frequency of RF field for the BSG6 sample as a function of DC magnetic field. (b) Typical FMR peak at 9 GHz for the BSG6 sample. The peak can be fitted by Lorentzian function to decompose the symmetric and antisymmetric components. (c) Extraction of saturation magnetization by the fit of resonant frequency as a function of resonant DC magnetic field. (d) Extraction of damping parameter by the fit of the width of EMF spectra as a function of resonant frequency.

## Note 4. Calculation of spin Hall angle in 3D system

For bulk SP-ISHE measurements, , $V_{sym}^{SCC}$ *vs.* thickness of NM layer ($d_N$) follows the relationship of $V_{sym}^{SCC} = I_c R = \theta_{SH} R w (\frac{2e}{\hbar}) \lambda \tanh(\frac{d_N}{2\lambda}) j_s^0$, where $\lambda$ is the spin diffusion length in the NM material.[2] To calculate spin Hall angle, one should know the value of spin diffusion length $\lambda$ in BSG. From the thickness dependence of $V_{sym}^{SCC}$, one can estimate the spin diffusion length in BSG should be smaller than 2 nm (see discussion part in main text). Therefore, the spin Hall angle can be obtained by injecting $\lambda = 2nm$, the extracted $j_s^0$ and the measured $V_{sym}^{SCC}$, $R$ as well as parameters of $w$, $d_N$ into the formula (S4):

$$\theta_{SH} = \frac{V_{sym}^{SCC}}{Rw(\frac{2e}{\hbar})\lambda \tanh(\frac{d_N}{2\lambda}) j_s^0} \quad \text{(S4)}$$

**Note 5. Renormalization THz signal with 5 nm Cu insertion**

As shown in Ref. [3], the detected THz signal intensity $E$ in FM/NM bilayer can be expressed as:

$$E(d) \propto \theta_{\mathrm{SH}} \lambda_{\mathrm{rel}} \frac{A}{d} \frac{\tanh(d_{\mathrm{N}}/2\lambda_{\mathrm{rel}})}{n_1+n_2+Z_0 \int_0^d dz\sigma(z)} \quad \text{(S5)}$$

Where $\theta_{\mathrm{SH}}$ is the spin-Hall angle in the NM layer. $\lambda_{\mathrm{rel}}$ can be considered as a hot-electron velocity relaxation length (typically 1 nm[2]). $A$ is the absorbed fraction of the incident pump power. $d_{\mathrm{N}}$ is the thickness of the NM layer. $d$ is the total thickness ($d_{\mathrm{F}}$+$d_{\mathrm{N}}$). $n_1$ is the refractive index of substrate. For MgO, sapphire or Si in THz domain, $n_1$=3.5. $n_2$ is the refractive index of air, $n_2$=1. $Z_0$ = 377 Ω is the vacuum impedance. $\sigma$ is the conductivity of metal layer.

For simplification, the THz signal intensity ratio γ between the BSG16 and BSG-Cu samples can be given by:

$$\gamma = \frac{n_1+n_2+Z_0(\sigma_{\mathrm{Cu}} d_{\mathrm{Cu}}+\sigma_{BSG} d_{BSG}+\sigma_{CFB} d_{\mathrm{CFB}})}{n_1+n_2+Z_0(\sigma_{BSG} d_{BSG}+\sigma_{CFB} d_{\mathrm{CFB}})} \quad \text{(S6)}$$

by injecting the parameters: $n_1$=3.5, $n_2$=1, $Z_0$ = 377 Ω, $\sigma_{\mathrm{Cu}}$=1/(20×10$^{-6}$) Ω$^{-1}$·cm$^{-1}$,[4] $\sigma_{BSG}$=1/(65×10$^{-3}$) Ω$^{-1}$·cm$^{-1}$, $\sigma_{CFB}$=1/(0.16×10$^{-3}$) Ω$^{-1}$·cm$^{-1}$, $d_{\mathrm{Cu}}$=5 nm, $d_{BSG}$=16 nm, $d_{\mathrm{CFB}}$=5 nm, we are able to obtain the ratio of 2.66.

**Note 6. XPS of BSG sample for ARPES measurement**

The $SiO_2$//MgO(2 nm)/BSG(8 nm) sample has been characterized by XPS in order to determine the chemical nature/ordering of the BSG layer being measured by ARPES. Since the sample was transferred through air into ARPES setup, we have carried out a soft $Ar^+$ etching on the surface to remove the oxidation. The results are presented in **Fig. S4**. On one hand, the evolution of O 1s core levels as function of cleaning (5 mins at 0.75 keV/cycle of $Ar^+$ etching) presented in **Fig. S4(a)** allows to identify the bulk (peak 1) and surface (peak 2) contributions. The bulk O signal could come from the bottom MgO layer due the limit mean free path of photoelectron. After two cycles of etching, the surface O component are much reduced. **Fig. S4(b)** presented the Bi 4f spectra after two cycles of etching as function of emission angle θ of the outgoing photoelectron. Two components with oxidized (high binding energy) and non-oxidized (low binding energy) states are evidenced. Increasing θ leads to a decrease of the oxidized component showing that the surface is nearly free from oxidation. However, deeper in the bulk, probably the BSG in contact with MgO is oxidized. On the other hand, Gd 3d (**Fig. S4(c)**) and Se 3d **(Fig. S4(d)**) show almost no oxidation signature confirming the high quality of the BSG samples. As a consequence, ARPES measurements being carried out at low energy (21.22 eV) are more sensitive to the surface of BSG layers than the BSG/MgO/$SiO_2$ interface.

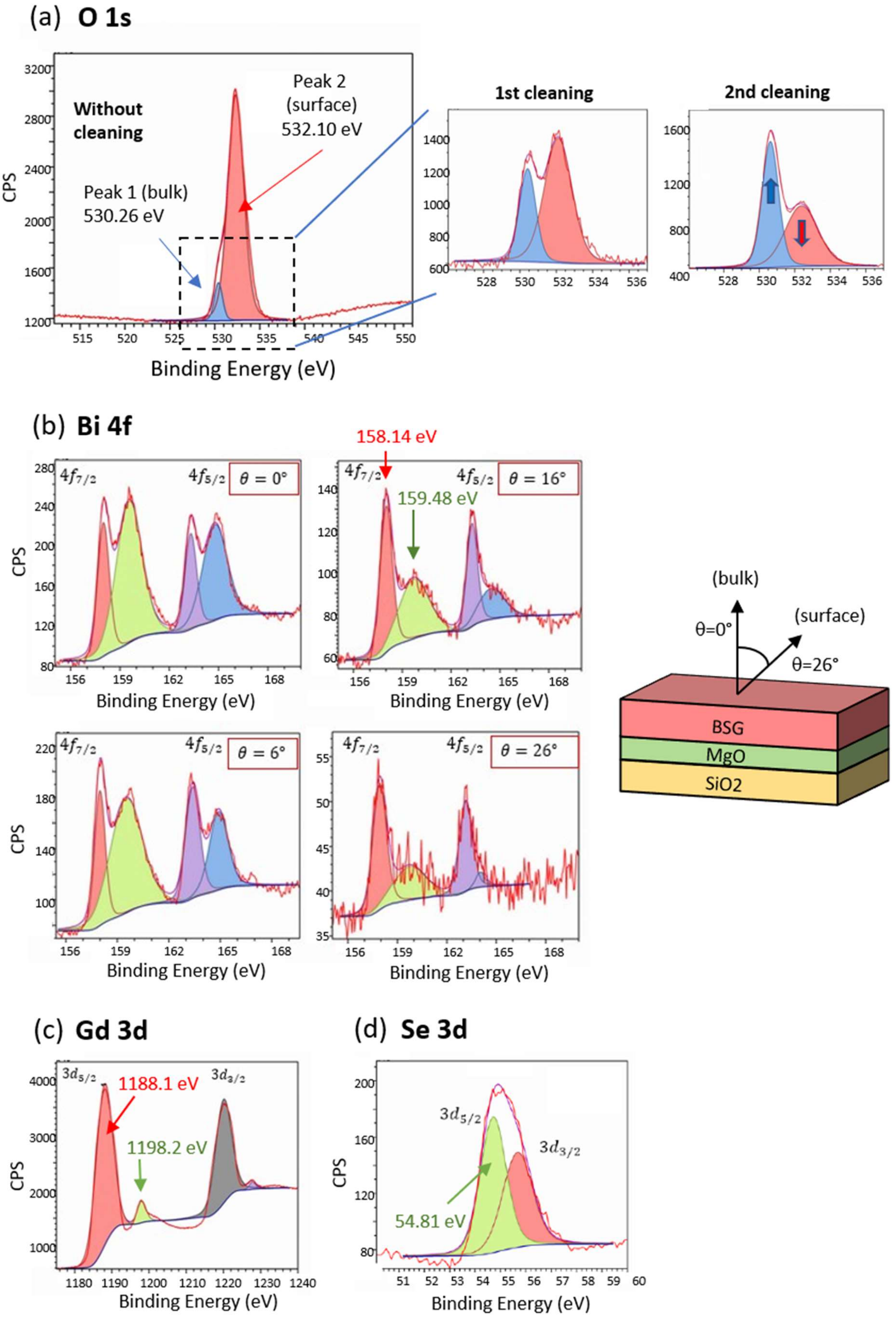

**Fig. S4:** (a) XPS spectra of O 1s before and after two cycles of $Ar^+$ etching cleaning. (b) XPS spectra of Bi 4f as function of emission angle θ of the outgoing photoelectron. Inset: schematics of detection angle with respect to the sample plane. (c) XPS spectra of Gd 3d. (d) XPS spectra of Se 3d.

**Note 7. AFM measurements of BSG layer with different thicknesses**

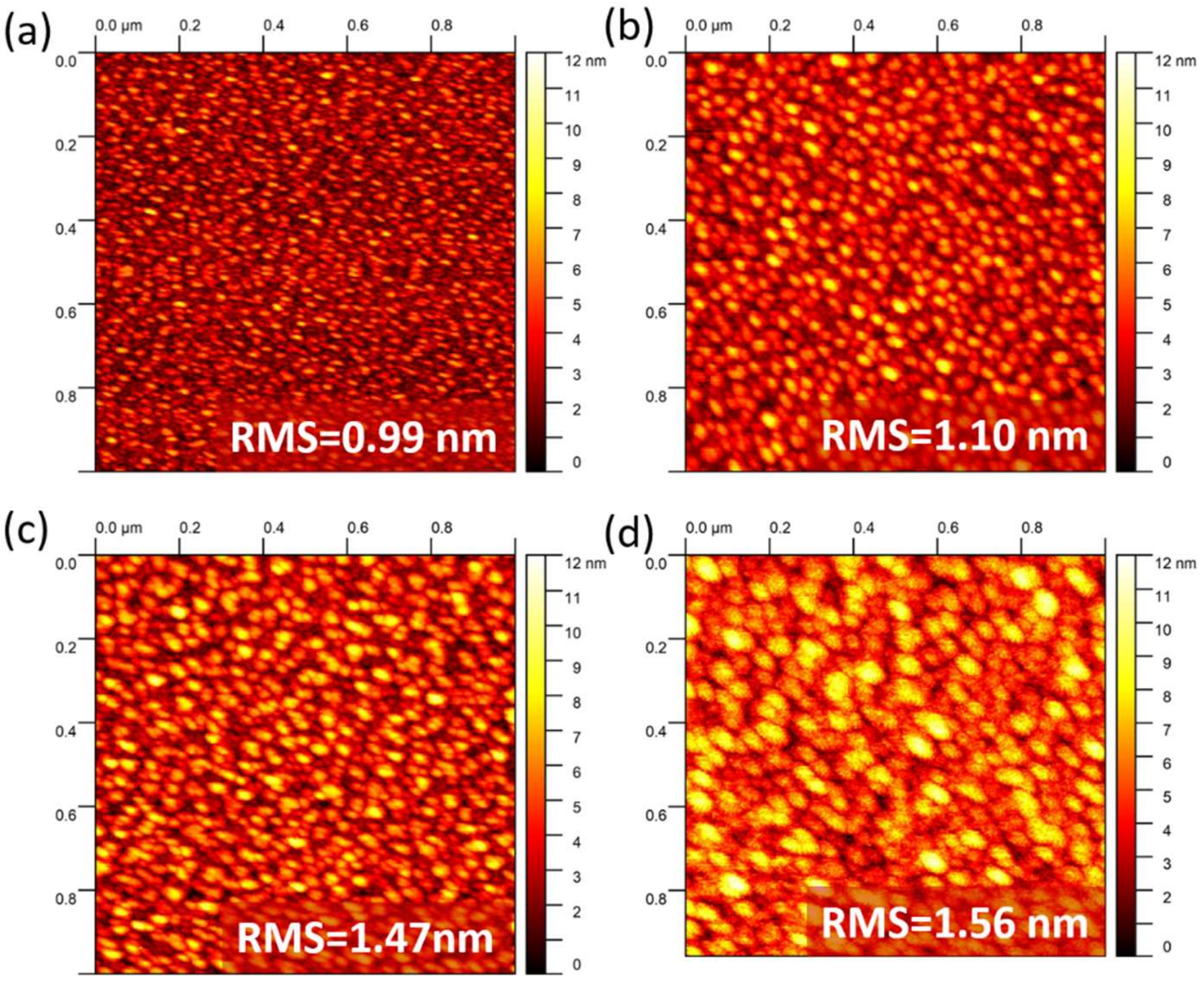

**Fig. S5:** AFM measurements of BSG layer with different thicknesses. (a) 6 nm, (b) 8 nm, (c) 12 nm, (d) 16 nm.

**Note 8. Comparison of sputtering grown topological insulator materials**

**Table S1:** Spin Hall angle (SHA), resistivity, characterization technique and morphology of sputtered topological insulators.

| Material | SHA | Resistivity (mΩ cm) | Characterization Technique | Morphology | Reference |
|---|---|---|---|---|---|
| $Bi_xSe_{1-x}$ | 18.6 | 12.8 | planar Hall | polycrystalline | 5 |
| $Bi_2Se_3$ | 2.76 | 3.9 | planar Hall | amorphous | 6 |
| $Bi_{0.85}Sb_{0.15}$ | 2.4 | 1.0 | harmonic Hall | polycrystalline | 7 |
| $Bi_{0.85}Sb_{0.15}$ | 10.7 | 0.7 | harmonic Hall | highly textured | 8 |
| $Bi_2Se_3$ | 0.11 | - | spin pumping | textured | 9 |
| BSG | 0.02 | 65 | spin pumping | amorphous | this work |